\documentclass[a4paper,11pt]{article}
\pdfoutput=1
\usepackage{jheppub}

\usepackage[utf8]{inputenc}

\usepackage{comment}
\usepackage{graphicx,rotating,amsmath,amsfonts,amssymb}
\usepackage{color}
\usepackage{enumitem}

\renewcommand{\Re}{\operatorname{Re}}
\renewcommand{\Im}{\operatorname{Im}}

\newcommand{\be}{\begin{equation}} 
\newcommand{\ee}{\end{equation}}

\def\lsim{\mathrel{\raise.3ex\hbox{$<$\kern-.75em\lower1ex\hbox{$\sim$}}}}
\def\gsim{\mathrel{\raise.3ex\hbox{$>$\kern-.75em\lower1ex\hbox{$\sim$}}}}

\newcommand{\tr}{\operatorname{tr}}

\newcommand{\td}{\mathrm{d}}

\newcommand{\eps}{\epsilon}
\newcommand{\inv}{\mathcal{I}}

\subheader{\small{Preprint: HIP-2017-19/TH}}
\title{Vacuum Stability and Perturbativity of SU(3) Scalars}

\author[a]{Matti Heikinheimo,}
\author[b]{Kristjan Kannike,}
\author[c]{Florian Lyonnet,}
\author[b]{Martti Raidal,}
\author[a]{Kimmo Tuominen,}
\author[b]{and Hardi Veerm\"ae}

\affiliation[a]{Helsinki Institute of Physics and Department of Physics, University of Helsinki
                      P.O.~Box 64, FI-00014, Helsinki, Finland}
\affiliation[b]{National Institute of Chemical Physics and Biophysics,  R\"avala 10, 10143 Tallinn, Estonia}
\affiliation[c]{Department of Physics, Southern Methodist University, 3215 Daniel Ave., Dallas, Texas, USA}

\emailAdd{matti.heikinheimo@helsinki.fi}
\emailAdd{kristjan.kannike@cern.ch}
\emailAdd{dibus2@gmail.com}
\emailAdd{martti.raidal@cern.ch}
\emailAdd{kimmo.i.tuominen@helsinki.fi}
\emailAdd{hardi.veermae@cern.ch}

\abstract{
We calculate the vacuum stability conditions and renormalisation group equations
for the extensions of standard model with a higher colour multiplet scalar
up to the representation $\mathbf{15'}$ that leaves the strong interaction asymptotically free.
In order to find the vacuum stability conditions, we calculate the orbit spaces for the self-couplings
of the higher multiplets, which for the representations $\mathbf{15}$ and $\mathbf{15'}$ of
$SU(3)_c$ are highly complicated. However, if the scalar potential is linear in orbit space variables,
it is sufficient to know the convex hull of the orbit space.
In contrast to the self-couplings of other multiplets, we find that
the scalar quartic couplings of the representations
$\mathbf{3}$ and $\mathbf{8}$ \emph{walk} rather than run, remaining nearly constant and perturbative
over a vast energy range. We describe the conditions for walking couplings using a schematic model.
With these technical results at hand we revise earlier results of generation of new scales with large $SU(3)_c$
scalar multiplets. Our results are easily extendable to models of new physics with additional $SU(3)$ or
SU($N$) gauge symmetries.
}

\begin{document}
\maketitle
\flushbottom

\section{Introduction}
\label{sec:in}

While the standard model (SM) fermion masses are protected by the chiral symmetry and, therefore, are predicted to be at or below the electroweak scale that is already probed by the LHC and
earlier colliders, the scalar content of Nature might be easily extended beyond the one presently known. An intriguing possibility is to consider scalar particles forming higher representations of the
QCD gauge group, the $SU(3)_c$. Higher multiplets confine at a higher scale~\cite{Marciano:1980zf,Lust:1985aw},
and one can entertain the idea that a scalar multiplet could condense at a new scale ${\cal O}(\rm{TeV})$ and trigger the electroweak symmetry breaking via the portal coupling to the SM Higgs
boson~\cite{Kubo:2014ova}. This mechanism could provide a dynamical and therefore natural explanation for the origin of the electroweak scale, akin to the framework of
technicolor~\cite{Weinberg:1979bn,Susskind:1978ms}.
Coloured scalar multiplets may also be utilised in builiding UV-complete models where complete asymptotic freedom~\cite{Pica:2016krb,Hansen:2017pwe}, i.e. a Gaussian UV fixed point of all couplings, arises.

There has also been an interest in the phenomenological signals of higher multiplets. The LHC signals have been studied for example for the sextuplet \cite{Berger:2010fy,Chen:2008hh} which can give rise to a diphoton signal \cite{Kats:2016kuz}. Both sextuplets \cite{Aydemir:2015oob} and octuplets \cite{Perez:2008ry,Dorsner:2009mq,Bertolini:2013vta} can arise in GUTs. In order to have Yukawa couplings with SM quarks, a sextuplet must have non-zero electric charge. Octuplets that are doublets of $SU(2)$ have been considered in the literature as well \cite{Manohar:2006ga,Cheng:2016tlc,Valencia:2016npc} with experimental constraints discussed e.g. in \cite{Dobrescu:2007yp,Gerbush:2007fe,Gresham:2007ri,Hayreter:2017wra}. An octuplet could also be seen via a di-Higgs signal \cite{Kribs:2012kz,Heng:2013cya,Nakamura:2017irk}. Decay of bound states of triplets, sextuplets and octuplets was discussed in \cite{Luo:2015yio}.
From the cosmological point of view the higher scalar representations of the QCD are useful to facilitate certain co-annihilation channels for dark matter (DM)~\cite{ElHedri:2016onc} that are absent for fundamental representation multiplets.  Another potential interest for higher representation scalars can be found in relation to the $CP$-symmetry and possible solutions to the strong $CP$-problem~\cite{Ratz:2016scn}.

While being theoretically motivated and interesting, the phenomenology of higher scalar representations of $SU(3)_c$ is technically challenging to handle. To overcome this difficulty, in this work we present a systematic study of all possible scalar colour multiplets that can be added to the SM particle content while preserving the asymptotic freedom of the QCD gauge coupling constant. For simplicity, we take the scalar multiplets to be neutral $SU(2)$ singlets.

We analyse the gauge orbit space of the scalar self-couplings to derive vacuum stability
constraints for the multiplets. Since the orbit space for the self-couplings is linear in the orbit space variables, it is sufficient to use the convex hull of the orbit space to find vacuum stability conditions. In addition, we calculate the renormalisation group equations (RGEs) for all the SM extensions under consideration and study the constraints on their parameter spaces from the perturbativity of the scalar quartic couplings.

We find that for most representations the scalar quartics develop Landau poles already far below the Planck scale, rendering the models for dynamical electroweak symmetry breaking via scalar colour multiplets susceptible to unknown nonperturbative effects. However, in certain cases, in particular for the representations $\mathbf{3}$ and $\mathbf{8}$, the quartic couplings of coloured scalars are very insensitive to radiative corrections and evolve very slowly, remaining almost constant over many decades of energy.
In this case the UV completion of the model is postponed many orders of magnitude over the Planck scale where gravity can be expected to influence the results in a crucial way.

We work out a generic description of these 
behaviours in terms of 
fixed points and show that this is applicable to several models, including the SM Higgs boson self-coupling. The results presented in this paper apply to coloured scalars as well as to models with a dark $SU(3)$ gauge group such as the $SU(3)$ dark matter~\cite{DiChiara:2015bua, Gross:2015cwa, Arcadi:2016kmk,Karam:2016rsz}.

The paper is organized as follows: In Section~\ref{sec:SM} we delimit the models we study from the requirement of asymptotic freedom. The  self-coupling potentials and vacuum stability conditions for higher-dimensional representations are derived in Section~\ref{sec:vacuum}. We study the running of quartic couplings in Section~\ref{sec:rge:higher} and draw our conclusions in Section~\ref{sec:out}. The full RGEs are presented in appendix~\ref{app:rge}, and in appendix~\ref{sec:appB} we give our bases for the representations $\mathbf{15}$ and $\mathbf{15'}$ of $SU(3)$.

\section{Higher multiplets and asymptotic freedom}
\label{sec:SM}

\begin{table}
\centering
\begin{tabular}{ccc}
Multiplet $\mathbf{R}$ & Casimir $C_2(\mathbf{R})$ & Index $T(\mathbf{R})$  \\
\hline
$\mathbf{3}$ & $\frac{4}{3}$ & $\frac{1}{2}$ \\
$\mathbf{6}$ & $\frac{10}{3}$ & $\frac{5}{2}$ \\
$\mathbf{8}$ & $3$ & $3$ \\
$\mathbf{10}$ & $6$ & $\frac{15}{2}$ \\
$\mathbf{15}$ & $\frac{16}{3}$ & $10$ \\
$\mathbf{15}^\prime$ & $\frac{28}{3}$ & $\frac{32}{2}$ \\
$\mathbf{21}$ & $\frac{40}{3}$ & $35$
\end{tabular}
\caption{The quadratic Casimir and the Dynkin index for higher multiplets of $SU(3)$.}
\label{table:reps}
\end{table}

For concreteness we consider the SM gauge group and particle content that is extended by one scalar multiplet charged under $SU(3)_c$.
In the choice of possible models, we require that after adding the new degrees of freedom, the theory should remain asymptotically free. The first coefficient of the $\beta$-function of the strong coupling $g_{3}$ is
\be
b_0=\frac{1}{12\pi}\left(\frac{11}{3}C_2(\mathbf{G})-\frac{4}{3}\sum_{\mathbf{R}_f}N_f T(\mathbf{R}_f)-\frac{1}{3}\sum_{\mathbf{R}_s}N_{s}T(\mathbf{R}_s)\right),
\ee
where $\mathbf{G}$, $\mathbf{R}_f$ and $\mathbf{R}_s$ indicate the gauge field, fermion and scalar representations, while $N_f$ and $N_s$ correspond to the number of Dirac fermions and complex scalars. For QCD we have $G=8$, and $N_f=6$ in the fundamental representation with $T(\mathbf{R}_f)=1/2$. Considering scalars in a single representation, the requirement $b_0>0$ implies
\be
N_s T(\mathbf{R}_s)< 33-2N_f=21.
\ee
From Table \ref{table:reps}, we find that for $N_s=1$ the representations $\mathbf{3}$, $\mathbf{6}$, $\mathbf{8}$, $\mathbf{10}$, $\mathbf{15}$ and $\mathbf{15'}$ of $SU(3)_{c}$ are allowed, while $\mathbf{21}$ breaks asymptotic freedom even when neglecting the contribution from the SM fermions. Consequently, we extend the SM particle content by one scalar $S$ in a higher multiplet $\mathbf{R}$ among these representations.

The Lagrangian reads
\be
\mathcal{L}=\mathcal{L}_{\rm SM}^{\mathrm{gauge, Yukawa}} - \mu_{H}^{2} |H|^{2} - m_{S}^{2} |S|^2- V_{\rm quartic},
\label{eq:V:quartic}
\ee
with the quartic part of the potential given by
\be
V_{\rm quartic}=\lambda |H|^2 + \lambda_{SH} |S|^2|H|^2 + V_{\bf R}(S),
\ee
where $V_{\bf R}(S)$ contains the self-interaction terms of $S$ in the representation $\mathbf{R}$ of $SU(3)_c$. The placement of indices of field tensors unambiguously distinguishes between the representation $\mathbf{R}$ and its complex conjugate $\mathbf{\bar{R}}$, so we will omit the bar or dagger from the latter. For example, for $S^{i}$ in $\mathbf{3}$, we have $S_{i} \equiv S^{i}{}^{\dagger}$. The potentials for the considered multiplets are given in Section~\ref{sec:vacuum} together with their vacuum stability conditions.

\section{Scalar potentials and conditions for the vacuum stability}
\label{sec:vacuum}

Any physical scalar potential must be bounded from below. In the limit of large field values, it suffices to study the quartic part of the potential \eqref{eq:V:quartic}. The self-coupling potential $V_{\mathbf{R}}(S)$ of $S$ can be written in terms of orbit space parameters $\rho_{i}$ associated with the representation $\mathbf{R}$ as
\be
	V_{\mathbf{R}} = (\lambda_{S} + \lambda_{Si} \rho_{i}) |S|^{4}.
	\label{eq:V:R}
\ee
The full potential \eqref{eq:V:quartic} is bounded from below if
\be
\label{eq:vac:stab:cond}
	\lambda_{H} > 0, \qquad
	\lambda_{S} + \lambda_{Si} \rho_{i} > 0, \qquad
	\lambda_{SH}  > -2 \sqrt{\lambda_{H} \left( \lambda_{S} + \lambda_{Si} \rho_{i} \right)},
\ee
for all allowed values of $\rho_{i}$. Therefore, the problem is reduced to determining the orbit space of $\rho_i$. The number of independent invariants $\rho_{i}$ for the representations considered here can be as large as 4 and thus the analysis can be quite involved. We will first discuss the mathematical structure of the vacuum stability conditions before moving on to specific examples.

\subsection{General considerations}

Consider first the general case of $N$ scalars $S_{a}$ without any reference to a particular gauge symmetry. The quartic term of the potential is then expressed as
\be
	V = \lambda_{abcd} S_{a}S_{b}S_{c}S_{d},
\ee
where summation over repeated indices is assumed. Symmetries will restrict the number of free parameters in the coupling tensor $\lambda_{abcd}$. It can be expressed as
\be
	\lambda_{abcd} = \lambda_{i} \hat\inv^{i}_{abcd},
\ee
where the index $i$ runs over all possible contractions allowed by the symmetries of the theory. The quartic potential can then be decomposed as
\be
	V = \lambda_{i}\, \inv_{i}		\qquad \text{with} \qquad
	\inv_{i} \equiv \hat\inv^{i}_{abcd}S_{a}S_{b}S_{c}S_{d}.
\ee
The invariants $\inv{i}$ are invariant under any symmetry transformations of the theory, in particular under gauge transformations.

The potential is bounded from below, if for all possible field values -- equivalently all values of $\inv_{i}$ -- we have
\be\label{eq:vac_cond_gen}
	V = \lambda_{i} \inv_{i} > 0.
\ee
There exists always at least one invariant $\inv_{0} = |S|^{4}$ that arises from the norm of the field. From the point of view of vacuum stability,  the element $\inv_{i}$ is equivalent to $\alpha \inv_{i}$ with $\alpha$ positive. This provides the space of $\inv_{i}$ naturally with the structure of a projective space. This projective space is the orbit space. To remove the redundancy, we can fix one of the \emph{non-negative} elements, for example require that $\inv_{0} = 1$. Equivalently one can work with the normalised orbit space parameters,
\be
	\rho_{i} = \frac{\inv_{i}}{\inv_{0}},
\ee
with $\rho_{0} = 1$, so, as in Eq.~\eqref{eq:V:R}, we can write
\be
  V = \lambda_{i} \rho_{i} |S|^{4},
\ee
and $V > 0$ if $\lambda_{i} \rho_{i} > 0$. In the rest of the paper we will represent the orbit space by the space of all possible configurations of $\rho_{i}$. It is straightforward to estimate the shape of the orbit space numerically by first determining a minimal set of independent invariants $\inv_{i}$, normalising them to obtain $\rho_{i}$, and then performing a numerical scan by evaluating the points $(\rho_{1},\rho_{2}, \ldots )$ for a large set of random field configurations $S_{a}$.

If the vacuum stability condition $\lambda_{i} \rho_{i} > 0$ is satisfied at two points $\rho_{i}^{A}$, $\rho_{i}^{B}$, then it is also satisfied at any point on a line connecting these two points:
\be\label{eq:vac_convexity}
	\lambda_{i} \rho_{i}^{A} > 0, \quad
	\lambda_{i} \rho_{i}^{B} > 0 \quad \implies \quad
	\lambda_{i} ( \eta \rho_{i}^{A} + (1-\eta)\rho_{i}^{B} ) > 0,
\ee
where $\eta \in [0,1]$ is an affine parameter.%
\footnote{The condition for non-normalised invariants would read $\eta_{A} \inv_{i}^{A} + \eta_{B} \inv_{i}^{B} > 0$ for any positive $\eta_{A}$, $\eta_{B}$. The condition $\inv_{0} = 1$ implies, however, that $\eta_{A} + \eta_{B} = 1$.}  If the potential depends linearly on the orbit space parameters, which is always the case for the self-couplings of a single multiplet, then Eq.~\eqref{eq:vac_convexity} implies that to determine whether the potential is bounded from below, one needs to know only the convex hull and not the exact shape of the orbit space itself.%
\footnote{That minima of the potential are likely to be on the cusps on the boundary of the orbit space was pointed out in \cite{Kim:1981xu} and later described in a qualitative way to lie on its convex hull in \cite{Degee:2012sk}.}

In the simplest case, the convex hull is a simplex. Because any point of the simplex can be given as a linear combination of its vertices, then by \eqref{eq:vac_convexity}, it is sufficient to require $\lambda_{i} \rho_{i} > 0$ not in all points of the orbit space, but only at the finite number vertices of its convex hull. If the convex hull of the orbit space is not a simplex, then it is always possible to approximate it with a simplex to an arbitrary accuracy, yielding a large number of approximate vacuum stability conditions. There exist well established numerical algorithms to find the convex hull of a set of points.

It is also possible, however, to find analytically at least a part of the orbit space, in particular its vertices.%
 The field configuration $S^{V}_{a}$ at a vertex has to satisfy \cite{Kim:1981jj}
\be\label{eq:cond:vertex}
	\frac{\partial \rho_i}{\partial S^{V}_{a}} = 0,
\ee
for all $i$ and $a$. This can be seen by considering a field configuration $S^{V}_{a}$ at a vertex $\rho^{V}_{i}$. By continuity in $\rho_{i}$, a small deviation of the field $S^{V}_{a} + \eps \delta S_{a}$ results generally in a small deviation $\rho^{V}_{i} + \eps\, \delta \rho_{i}$, where $\eps$ is an infinitesimal quantity. On the other hand, this deviation can not move the point out of the orbit space. Because the vertex is an object of dimension zero, the only way to guarantee it, is to require $\delta \rho_{i} = 0$.
The vertices of the orbit space may or may not coincide with the vertices of its convex hull. For example, there may be vertices in \emph{concave} parts of the orbit space, i.e.  sticking in, not out.

Based on the similar reasoning it is possible to study the edges, faces, and in general, $k$-faces of non-simplical convex hulls. To this purpose consider the rank of the matrix
\be
	J_{ia} = \frac{\partial \rho_i}{\partial S_{a}}.
	\label{eq:face:rank}
\ee
Its value corresponds to the dimensionality of the $k$-face of the convex hull, e.g.  if the rank of $J_{ia}$ is $0$, then the solution $S_{a}$ gives a vertex, and $\mathrm{rank}(J_{ia})  = 1$  corresponds to an edge. This is a necessary, but not sufficient condition. It is possible to simplify \eqref{eq:face:rank} by using gauge rotations to set a number of field components to zero.

\subsection{Vacuum stability for the representations $\mathbf{3}$ and $\mathbf{8}$}

In case of the representations $\mathbf{3}$ and $\mathbf{8}$ only one self-interaction term exists.  In detail, for the $S^{i}$ in the representation $\mathbf{3}$ of $SU(3)$ the potential reads
\be\label{eq:V:3}
  V_{\bf 3}(S) = \lambda_{S} (S_{i} S^{i})^{2},
\ee
where $S_{i} \equiv S^{\dagger i}$. The representation $\mathbf{8}$ of $SU(3)$ consists of Hermitian and traceless matrices $S^{i}_{j}$ and the general quartic term is%
\footnote{Note that since $\mathbf{8}$ is Hermitian and \emph{traceless}, singlet under $SU(2)_{L}$ and neutral, its self-interaction potential contains a single term. In general, there would be several more terms, as in \cite{Manohar:2006ga}, for example.}
\be
  V_{\bf 8}(S) = \lambda_{S} (\tr{S^{2}})^{2}.
  \label{eq:V:8}
\ee
Both orbit spaces therefore contain a single point and thus the vacuum stability condition for the self-interaction potential \eqref{eq:V:8} simply reads
\be
	\lambda_{S} > 0.
\ee

\subsection{Vacuum stability for the representations $\mathbf{6}$ and $\mathbf{10}$}

Both the representations $\mathbf{6}$ and $\mathbf{10}$ have two independent quartic terms. The representation $\mathbf{6}$ is given by the symmetric matrix $S^{ij}$. The self-interaction potential for the $S^{i}$ in the $\mathbf{6}$ of $SU(3)$ is
\be
	V_{\bf 6}(S)
	= \lambda_{S} (\tr{S^{\dagger} S})^{2} + \lambda_{S1} \tr{S^{\dagger} S S^{\dagger} S}.
\ee
The matrix $S$ can be diagonalised inside the traces, so the orbit parameter has the form
\be
	\rho = \frac{\tr{S^{\dagger} S S^{\dagger} S}}{(\tr{S^{\dagger} S})^{2}} = \frac{\sum_{i} |d_{i}|^{4} }{(\sum_{i} |d_{i}|^{2})^{2}},
	\label{eq:rho:6}
\ee
where the summation runs over the three eigenvalues $d_{i}$ of the matrix $S$. It is now straightforward to demonstrate that $\rho_{\text{min}} = 1/3$ and $\rho_{\text{max}} = 1$. The minimum is obtained for all $d_{i}$ equal, the maximum is obtained if one of them is non-zero and the rest are zero. The orbit space is therefore the interval $\rho \in [1/3,1]$, which is a one-dimensional simplex, and the sufficient and necessary conditions for stability of the vacuum correspond to its endpoints:
\be\label{eq:vac_stab:6}
	\lambda_{S} + \lambda_{S1} > 0, \qquad
	\lambda_{S} + \frac{1}{3}\lambda_{S1} > 0.
\ee

Likewise, the potential for the representation $\mathbf{10}$ can be written in terms of the positive symmetric matrix $M_{i}^{j} = S_{ikl} S^{jkl}$ as
\be
	V_{\mathbf{10}} = \lambda_{S} (\tr{ M })^{2} + \lambda_{S1} \tr M^{2}.
\ee
By the same argument, also here $\rho_{\text{min}} = 1/3$ and $\rho_{\text{max}} = 1$ and the vacuum stability condtitions are identical to \eqref{eq:vac_stab:6}.

\subsection{Vacuum stability for the representation $\mathbf{15'}$}

\begin{figure}[tb]
\begin{center}
\includegraphics{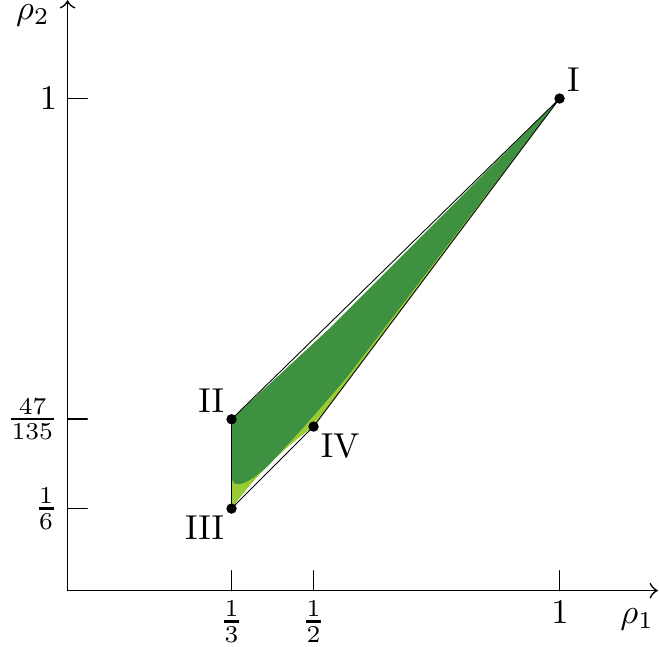}
\caption{The orbit space of the representation $\mathbf{15'}$ is shown in green. Dark green represents the region where a $U(1)$ subgroup can be preserved. An $SU(2)$ subgroup is only preserved on the cusp $\rho_{1} = \rho_{2} = 1$. The convex hull of the orbit space is a simplex. Its borders are shown with lines joining the vertices of the orbit space.}
\label{fig:orbit:space:15:p}
\end{center}
\end{figure}

The representation $\mathbf{15'}$ forms a completely symmetric 4th order tensor $S^{ijkl}$. Its self-interaction potential is given by
\be
  V_{\bf 15'}(S) = \lambda_{S} (S_{ijkl} S^{ijkl})^{2} + \lambda_{S1} S_{ijkp} S^{ijkq} S_{lmnq} S^{lmnp}
  + \lambda_{S2} S_{ijmn} S^{ijpq} S_{klpq} S^{klmn}.
  \label{eq:V:15'}
\ee
The ranges of the two orbit space parameters can be understood if we recast the potential \eqref{eq:V:15'} as
\be
	V_{\bf 15'}(S) = \lambda_{S} (\tr M)^{2} + \lambda_{S1}  \tr M^{2} + \lambda_{S2} \tr M'^{2},
\ee
where $M^{p}_{q} \equiv S_{ijkp} S^{ijkq}$ and $M'^{mn}_{pq} \equiv S_{ijmn} S^{ijpq}$. The combination $M^{p}_{q}$ is a $3\times3$ matrix and $M'^{mn}_{pq}$ can be treated as a multi-index $6\times6$ matrix. Note that $\tr M = \tr M'$. Therefore
\be\label{eq:15p_rho}
  \rho_{1} = \frac{\tr M^{2}}{(\tr M)^{2}}, \qquad
  \rho_{2} = \frac{\tr M'^{2}}{(\tr M)^{2}},
\ee
and by the argument of the previous subsection, the two orbit parameters are bounded by $1/3 \leqslant \rho_1 \leqslant 1$ and $1/6 \leqslant \rho_{2} \leqslant 1$. Alas, since the elements of the matrices $M$ and $M'$ have non-trivial dependencies due to their construction from $S^{ijkl}$, they do not provide an easy way to find the true shape of the orbit space within the rectangle defined by these inequalities.

Using the condition \eqref{eq:cond:vertex} we have determined that the orbit space has four vertices at the points $\vec\rho_{\mathrm{I}} = (1, 1)$, $\vec\rho_{\mathrm{II}} = (1/3, 47/135)$, $\vec\rho_{\mathrm{III}} = (1/3, 1/6)$ and $\vec\rho_{\mathrm{IV}} = (1/2, 1/3)$. The convex hull of the orbit space coincides with the simplex built from these four points. In conclusion, we obtain the following necessary and sufficient vacuum stability conditions:
\be
\begin{split}
	\text{I}\,: 	\qquad&  \lambda_{S} + \lambda_{S1} + \lambda_{S2} > 0,\\
	\text{II}\,: 	\qquad&  \lambda_{S} + \frac{1}{3}\lambda_{S1} + \frac{47}{135}\lambda_{S2} > 0,\\
	\text{III}\,: 	\qquad&  \lambda_{S} + \frac{1}{3}\lambda_{S1} + \frac{1}{6}\lambda_{S2} > 0, \\
	\text{IV}\,: 	\qquad&  \lambda_{S} + \frac{1}{2}\lambda_{S1} + \frac{1}{3}\lambda_{S2} > 0.
\end{split}
\ee
The orbit space and its convex hull are depicted in Fig.~\ref{fig:orbit:space:15:p}.  The orbit space was evaluated by a numerical scan over random field configurations. We found that the orbit space can be covered by considering only a subset of possible field configurations given by non-vanishing values for the parameters $a_{1}$, $a_{5}$, $a_{11}$, $a_{13}$, $a_{15}$ as defined in Appendix~\ref{sec:param:15p}. As a consistency check we have numerically tested whether the points $\vec\rho_{\rm I - IV}$ depicted in Fig.~\ref{fig:orbit:space:15:p} determine the convex hull of the orbit space, by evaluating extremal slopes of lines passing through a given cusp  $\vec\rho_{\rm I}$, $\vec\rho_{\rm II} $, $\vec\rho_{\rm III}$ or $\vec\rho_{\rm IV}$ and an arbitrary point of the orbit space.\footnote{This amounts to finding the extrema of the quantity $(\rho_1 - \rho^*_1)(\rho_2 - \rho^*_2)$, where $(\rho^*_1,\rho^*_2)$ is a fixed cusp $\vec\rho_{\rm I}$, $\vec\rho_{\rm II} $, $\vec\rho_{\rm III}$ or $\vec\rho_{\rm IV}$ and $(\rho_1,\rho_2)$ is a function of the field.}

It is interesting to study how the orbit space relates to symmetry breaking \cite{Abud:1981tf,Abud:1983id,Kim:1981xu,Kim:1983mc}. For example, if a $SU(3)$ fundamental (triplet) scalar should obtain a vev, the $SU(3)$ gauge symmetry is always broken to $SU(2)$. This can be seen by considering the action of the $SU(3)$ generators on the triplet and noting that the three generators whose action vanishes obey the Lie algebra of $SU(2)$. This method can be easily extended to higher representations, that allow for more complicated symmetry breaking patterns. Up to gauge transformations,  for $\mathbf{15'}$ the only vev invariant under $SU(2)$ is given by $S_{1111} \neq 0$ and all other elements zero. Plugging this field configuration into the expressions of the orbit space parameters \eqref{eq:15p_rho}, we find that the breaking pattern $SU(3) \to SU(2)$ corresponds to a single point of the orbit space: the cusp $\vec\rho_{\rm I} = (1,1)$. 
Field configurations symmetric under a $U(1)$ subgroup populate the dark green area in Fig.~\ref{fig:orbit:space:15:p}, and any field configuration corresponding to the light green area in Fig.~\ref{fig:orbit:space:15:p} will break $SU(3)$ completely.

\subsection{Vacuum stability for the representation $\mathbf{15}$}

The general element of the representation $\mathbf{15}$ of $SU(3)$ is given by a tensor $S_{k}^{ij}$ that is traceless and symmetric in the upper indices. The self-interaction potential,
\be
\begin{split}
  V_{\bf 15}(S)
  &= \lambda_{S} (S_{ij}^{k} S_{k}^{ij})^{2} + \lambda_{S1} S_{jm}^{i} S_{i}^{jn} S_{ln}^{k} S_{k}^{lm}
  + \lambda_{S2} S_{jm}^{i} S_{i}^{jn} S_{kl}^{m} S_{n}^{kl}
 \\
  &+ \lambda_{S3} S_{ij}^{m} S_{n}^{ij} S_{kl}^{n} S_{m}^{kl} + \lambda_{S4} S_{jm}^{i} S_{l}^{km} S_{in}^{j} S_{k}^{ln},
\end{split}
\ee
contains five independent terms and the orbit space is therefore four-dimensional. The orbit parameters are bounded by
\be\label{eq:bounds_15p}
	\frac{1}{3} \leqslant \rho_{1} \leqslant 1, \qquad
	0 \leqslant \rho_{2} \leqslant \frac{1}{2}, \qquad
	\frac{1}{3} \leqslant \rho_{3} \leqslant 1, \qquad
	0 \leqslant \rho_{4} \leqslant \frac{9}{16}.
\ee
The ranges of $\rho_{1}$ and $\rho_{3}$ can be understood by writing them in terms of the matrices $M_{m}^{\prime n} = S_{jm}^{i} S_{i}^{jn}$ and $M_{m}^{\prime \prime n} = S_{ij}^{n} S_{m}^{ij}$ with $\tr M' = \tr M'' = S_{ij}^{k} S_{k}^{ij}$. Then
\be
	\rho_{1} = \frac{\tr M^{\prime 2}}{(\tr M')^{2}}, \qquad \rho_{3} = \frac{\tr M^{\prime\prime 2}}{(\tr M'')^{2}},
\ee
and we can apply the argument of eq. \eqref{eq:rho:6} yet again.
 In addition,
\be
	\rho_{2} = \frac{\tr M' M''}{\tr M' \tr M''},
\ee
which yields the minimal value $\rho_{2} = 0$ if $M'$ and $M''$ are orthogonal.

We obtain the minimum for $\rho_{2} = \rho_{4} = 0$, e.g., for $a_{1} = 1$ and all other $a_{i} = 0$ (see Appendix~\ref{sec:param:15} for our basis for the $\mathbf{15}$). The maximum value of $\rho_{2} = 1/2$ is obtained for $a_{12} = a_{7}$ and all other $a_{i} = 0$, for example. The maximum value of $\rho_{4} = 9/16$ instead of unity is due to the tracelessness conditions on $S^{ij}_{k}$ (it is obtained for $a_{11} \neq 0$ and all other $a_{i} = 0$).

As before, the shape of the orbit space is much more complicated than indicated by the inequalities \eqref{eq:bounds_15p}. Table~\ref{tab:15:vertices} lists the six vertices of the orbit space. The vertices were obtained by a process of educated trial and error. They can by found by using eq. \eqref{eq:cond:vertex} together with taking only one or a few elements non-zero at a time. A minimal set of elements that yields all the vertices is given by $a_{1}$, $a_{7}$, $a_{10}$ and $a_{11}$. We checked that adding one or two additional elements at a time did not produce additional vertices.%
\footnote{Note that an $SU(3)$ transformation allows to set seven elements to zero so the maximum number of elements needed to produce the vertices cannot surpass eight.}

\begin{figure}[tb]
\begin{center}
  \includegraphics{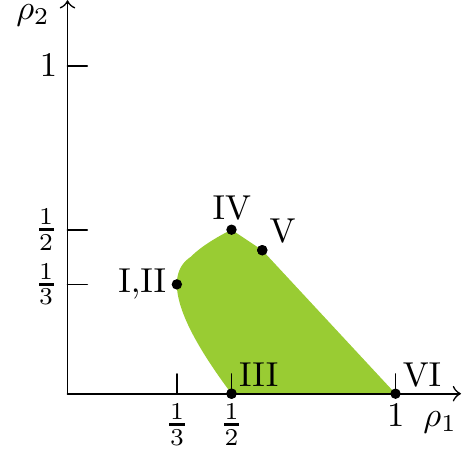}
  \includegraphics{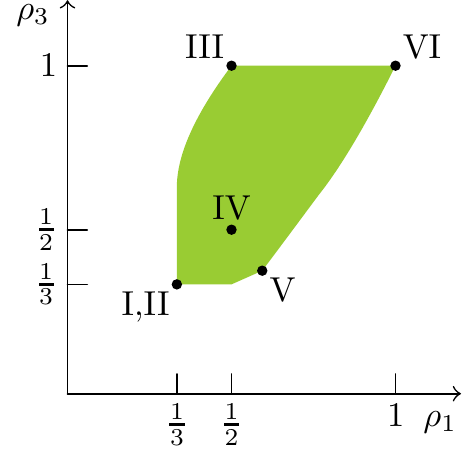}
  \includegraphics{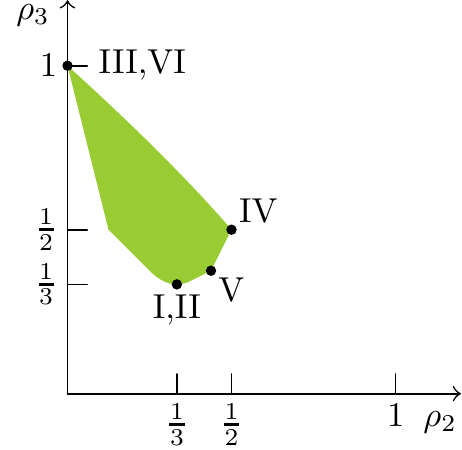}
    \\
  \includegraphics{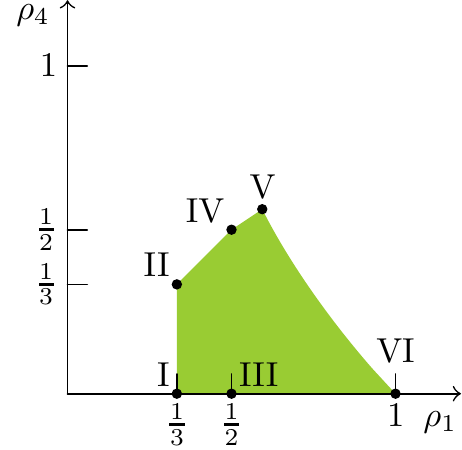}
  \includegraphics{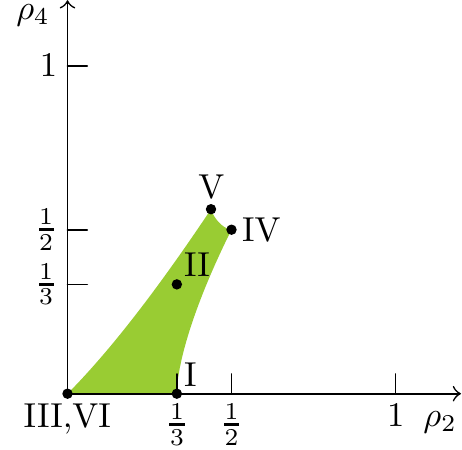}
  \includegraphics{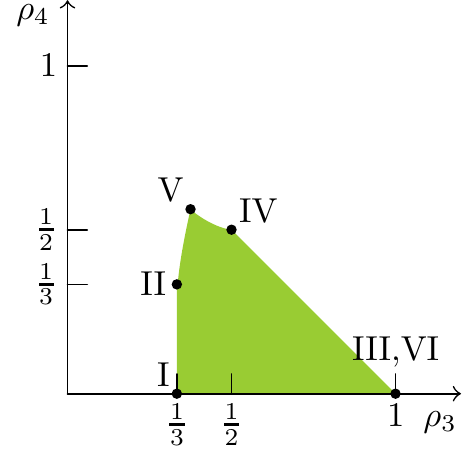}
\caption{Projections of the four-dimensional orbit space of the representation $\mathbf{15}$.}
\label{fig:orbit:space:15}
\end{center}
\end{figure}

Fig.~\ref{fig:orbit:space:15} shows two-dimensional projections of the four-dimensional orbit space obtained by a random scan over field configurations. In this case, because the orbit space has convex curved parts, the convex hull of the orbit space is not a simplex and therefore the vertices fail to provide sufficient vacuum stability conditions. Insertion of the vertices listed in Table~\ref{tab:15:vertices} into  Eq.~\eqref{eq:vac:stab:cond} yields six \emph{necessary} vacuum stability conditions.

If the couplings obey the necessary requirements but fail to satisfy the sufficient criteria, vacuum stability has to be checked on the whole orbit space. Though it would be rather complicated to find its analytical form, the convex hull of the orbit space can be established via numerical computation to a high precision. We use Loren Petrich's Mathematica code \href{http://lpetrich.org/Science/\#CHDV}{\texttt{http://lpetrich.org/Science/\#CHDV}} to find the four-dimensional convex hull. We include the points of the 4-dimensional convex hull of the orbit space as an ancillary file \texttt{ch15.dat} with the \LaTeX\ source of the paper.

\begin{table}[tb]
\caption{Vertices of the orbit space of the self-couplings of $S$ in the $\mathbf{15}$ of $SU(3)$.}
\begin{center}
\begin{tabular}{ccccc}
Vertex & $\rho_{1}$ & $\rho_{2}$ & $\rho_{3}$ & $\rho_{4}$
\\
\hline
I & $\frac{1}{3}$ & $\frac{1}{3}$ & $\frac{1}{3}$ & $0$ \\
II & $\frac{1}{3}$ & $\frac{1}{3}$ & $\frac{1}{3}$ & $\frac{1}{3}$ \\
III & $\frac{1}{2}$ & $0$ & $1$ & $0$ \\
IV & $\frac{1}{2}$ & $\frac{1}{2}$ & $\frac{1}{2}$ & $\frac{1}{2}$ \\
V & $\frac{19}{32}$ & $\frac{7}{16}$ & $\frac{3}{8}$ & $\frac{9}{16}$ \\
VI & $1$ & $0$ & $1$ & $0$
\end{tabular}
\end{center}
\label{tab:15:vertices}
\end{table}%

\section{RGE running of higher $SU(3)_{c}$ multiplets}
\label{sec:rge:higher}

In this section we study the running of coupling constants for higher $SU(3)_{c}$ multiplets. We find that the quartic couplings have a Landau pole for all representations. It is possible, however, to construct viable scenarios for which this Landau pole appears at scales far above the Planck scale. We will refer to the tendency of a coupling to remain constant and perturbative for a vast energy range as {\it walking}\footnote{Similar terminology is used in the context of walking technicolor to characterise the evolution of the gauge coupling.} (as opposed to running).

\subsection{Walking in a schematic model}
\label{sec:toy:model}

To better understand the concept of walking and the circumstances in which it can occur, consider first an exactly solvable model consisting of a scalar multiplet with non-Abelian interactions and a single quartic term. At one-loop level, its RGEs have the general form
\be\label{ex:RGE_walking}
	\frac{\td g^{2}}{\td t} = - b_{g} g^{4}, \qquad
	\frac{\td \lambda}{\td t} =  b_{\lambda} \lambda^{2} - b_{\lambda g} g^{2} \lambda   + b_{\lambda gg} g^{4},
\ee
where $t = \ln  \mu$ and the coefficients are determined by the details of the theory. In the following analysis we assume $b_{g}, b_{\lambda} > 0$. The RGEs are invariant under a shift symmetry $t \to t + \Delta t$ and under the scaling symmetry
\be\label{eq:general_1L_RGE_sym}
	g_i(t) \to \sqrt{c}\,g_i(c t), 	\qquad
	\lambda_i(t) \to c\,\lambda_i(c t),
\ee
where $c$ and $\Delta t$ are arbitrary constants. We first find a particular solution and then use the fact that all other solutions can be obtained by making use of the symmetries \eqref{eq:general_1L_RGE_sym}.
The first equation in Eq.~\eqref{ex:RGE_walking} is solved by
\be\label{ex:sol_g_0}
	g^{2} = \frac{1}{b_{g} t}.
\ee
Inserting it into the second equation, we find
\be\label{ex:RGE_walking_2}
	\frac{\td \bar\lambda}{\td \ln\,t}
	= b_{\lambda} \bar \lambda^{2} - \frac{b_{\lambda g}}{b_{g}} \bar \lambda   + \frac{b_{\lambda gg}}{b_{g}^{2}} \\
	\equiv b_{\lambda} (\bar\lambda - \bar\lambda_{+})(\bar\lambda - \bar\lambda_{-}),
\ee
where $\bar \lambda \equiv t \lambda$ and $\bar\lambda_{\pm}$ are the roots of the polynomial on the RHS of Eq.~\eqref{ex:RGE_walking_2}.  Real $\bar\lambda_{\pm}$ are fixed points of the system and we order them as $\bar\lambda_{+} \geqslant \bar\lambda_{-}$ so that $\bar\lambda_{+}$ represents an UV repellor and $\bar\lambda_{-}$ is a UV attractor.\footnote{Linear stability of \eqref{ex:RGE_walking_2} at $\bar\lambda_{-}$ implies that $b_{\lambda}(\bar\lambda_{+} - \bar\lambda_{-}) \equiv 2\delta > 0$.}
Vacuum stability additionally demands $\bar\lambda_{-} \geqslant 0$ from a physical attractor.  Note, however, that the roots $\bar\lambda_{\pm}$ could also be complex.   A particular solution of \eqref{ex:RGE_walking_2} reads
\be\label{ex:sol_lambda_0}
	\bar\lambda = \frac{1}{b_{\lambda}} \, \left[\sigma  - \frac{\delta}{\tanh(\delta \ln t )}\right],
\ee
where
\be\label{eq:sigma:delta}
	\sigma \equiv \frac{b_{\lambda}}{2} (\bar\lambda_{+} + \bar\lambda_{-})
	= \frac{1}{2}\left(\frac{b_{\lambda g}}{b_{g}} - 1\right), \qquad
	\delta  \equiv \frac{b_{\lambda}}{2} (\bar\lambda_{+} - \bar\lambda_{-})
	= \sqrt{\sigma^{2} - \frac{b_{\lambda}b_{\lambda gg}}{b_{g}^{2}} },
\ee
denote the rescaled half-difference and average of the roots.

\begin{figure}[tb]
\begin{center}
\includegraphics[width=7cm]{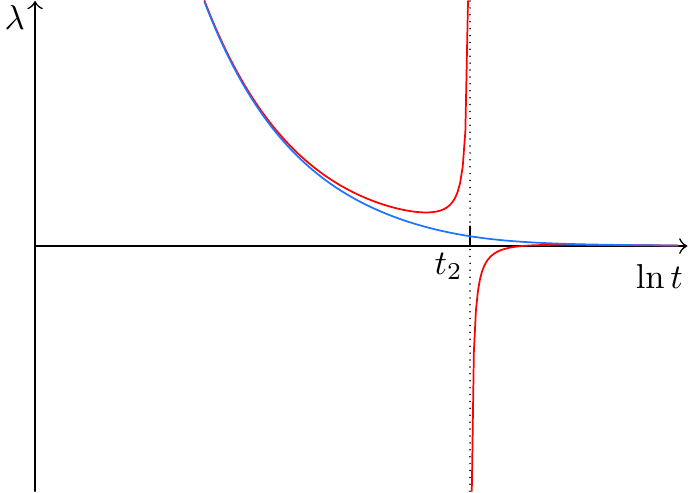}~~
\includegraphics[width=7cm]{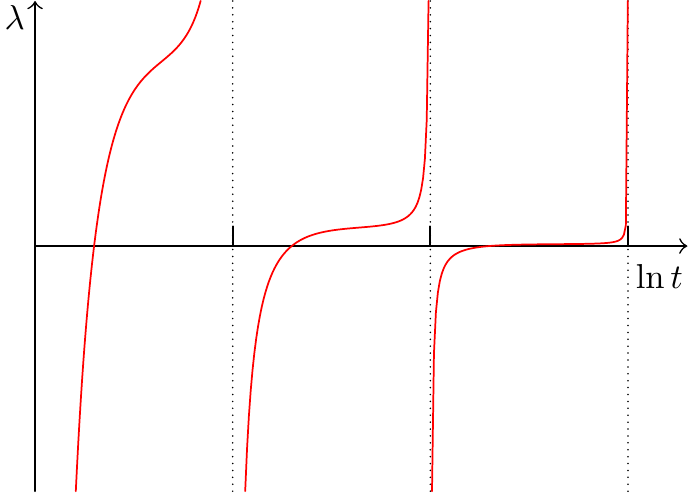}
\caption{Solutions for the scalar quartic $\lambda$ in the model~\eqref{ex:RGE_walking}. The IR pole of $g^{2}$ lies at the coordinate origin, i.e. $t_{2} = 0$. The positions of other poles of $\lambda$ are depicted by the vertical dotted lines. Left: Real $\delta$. The first family of solutions is shown in blue, the second one in red. Right: Imaginary $\delta$. These solutions can exhibit walking behaviour.}
\label{fig:toy:model:run}
\end{center}
\end{figure}

The general solution can now be obtained
\be\label{ex:toy_sol_gen}
	g^{2} = \frac{1}{b_{g} (t - t_{1})}, \qquad
	\lambda = \frac{1}{b_{\lambda}(t - t_{1})} \, \left[ \sigma  - \frac{\delta}{\tanh\left( \delta \ln \frac{t - t_{1}}{t_{2} - t_{1}} \right)}\right],
\ee
where $t_{1}$ and $t_{2}$ are constants of integration that were inserted by applying the scaling symmetry~\eqref{eq:general_1L_RGE_sym} and shift symmetry. They can be determined from initial conditions, that is, from $g(\mu_0)^2 = g_0^2$ and $\lambda(\mu_0) = \lambda_0$, where $g_0$ and $\lambda_0$ are the values of the couplings measured at the scale $\mu_0$. Reality of the solutions does not forbid a complex $t_{2}$ but it implies that $\Re t_{2} > t_{1}$. All solutions are singular at $t = t_{1}$ i.e.  at the IR pole of gauge coupling. Additional singularities arise from zeroes of the hyperbolic arctangent. There are two qualitatively different cases:
\begin{enumerate}[leftmargin=*]
	\item For a real $\delta$ there are two families of real solutions connected by rescalings \eqref{eq:general_1L_RGE_sym} with a complex $c$.

	First, if $\Im (\ln (t_{2} - t_{1}) = 0)$, then there is an additional singularity at $t_{2}$. This solution is depicted by a red line in the left panel of Fig.~\ref{fig:toy:model:run}. The running is asymptotically free for $t >  t_{2}$ or equivalently if $\bar\lambda< \bar\lambda_{-} $. This case is interesting because the vacuum is stable if $\bar\lambda_{-} > 0$ but the theory has a pole in the IR which indicates that the self interaction of the scalars becomes strong before confinement, because $t_{2} > t_{1}$.
  The branch at $t < t_{2}$, corresponding to $\bar\lambda < \bar\lambda_{+}$, has a UV pole signalling the breakdown of our one loop analysis.

	The second family is given by the blue line in the left panel of Fig.~\ref{fig:toy:model:run}. It obeys $\Im (\ln (t_{2} - t_{1})) = \pi/(2\delta)$ which effectively corresponds to taking the reciprocal of the hyperbolic arctangent. There are no additional singularities, $\bar\lambda$ varies in the interval $(\bar\lambda_{-}, \bar\lambda_{+})$ and all solutions are asymptotic to $\bar\lambda_{-}/t$ in the UV. The model exhibits total asymptotic freedom.

	\item The case of imaginary $\delta$ is shown in the right panel of Fig.~\ref{fig:toy:model:run}. On top of $t_{1}$, $\lambda$ is singular at infinite set of points given by
\be\label{2-poles}
	t_{\Lambda, n} = t_{1} + (t_{2} - t_{1})e^{n \pi/|\delta|} >  t_{1},
\ee
	with $n$ an integer. Perturbative physics -- consistent with our one-loop analysis -- always lies between two poles and thus there are a infinite number of families of real solutions. The inequality $t_{\Lambda, n} > t_{1}$, follows from $t_{2} > t_{1}$ and guarantees that the quartic always hits its IR pole before the gauge coupling, implying that, running towards lower energies, the self-interaction of the scalars becomes strong before the gauge interaction.

	\item In the extremal case $\delta = 0$ the general solution \eqref{ex:toy_sol_gen} reads
\be\label{ex:toy_sol_gen_d0}
	g^{2} = \frac{1}{b_{g} (t - t_{1})}, \qquad
	\lambda = \frac{1}{b_{\lambda}(t - t_{1})} \, \left[ \sigma  - \frac{1}{\ln \frac{t - t_{1}}{t_{2} - t_{1}}}\right],
\ee
and the singularity at $t_{2}$ is approached as a double logarithm of the energy scale.
\end{enumerate}

For real $\delta$ the solutions could be determined by knowing the poles $t_{1}$ and $t_{2}$, the distance between poles has a one to one correspondence with the initial conditions. For imaginary $\delta$ this is not the case, as on top of $t_{1}$ and $t_{2}$ on needs the specify the family given by the integer $n$. The poles of families with higher $n$ will be exponentially separated whenever $|\delta| \ll \pi$.

We now limit our discussion to the cases that do not support total asymptotic freedom, that is $\Re\delta = 0$, and estimate the energy range at which the theory can remain perturbative. We assume that the scalar has a mass $m_{S}$ and that $\lambda(m_{S}) \approx 0$. The mass scale lies between two poles $t_{\Lambda, n-1} < \ln(m_{S}/\mu) < t_{\Lambda, n} \equiv \ln(\Lambda/\mu)$, where $\Lambda$ denotes the UV pole of $\lambda$ and $\mu$ is an arbitrary reference scale introduced for dimensional reasons. At scales $\mu<m_{S}$ the scalar decouples and the IR behavior of the quartic coupling becomes irrelevant.

From Eq.~\eqref{2-poles} we obtain the upper bound on the scale of the Landau pole
\be\label{toy_bound}
	\ln(\Lambda/m_{S}) \leqslant \ln(m_{S}/\Lambda_{g}) \left( e^{\pi/|\delta|} - 1 \right),
\ee
where $\Lambda_{g}$, defined by $\ln(\Lambda_{g}/\mu) \equiv t_{1}$, denotes the confinement scale of the gauge interaction. In case $|\delta| \gg \pi$ the energy range where the theory can be non-perturbative is unavoidably small. A small $|\delta|$, on the other hand, can naturally accommodate a large separation between the Landau pole and the mass scale. This is characteristic to walking.

A similar walking behaviour can be inferred from the $\delta = 0$ solution \eqref{ex:toy_sol_gen_d0}, that contains a Landau pole if $b_{\lambda} > 0$. Although the separation between the IR and UV poles is determined through the initial conditions that fix $t_{1}-t_{2} \equiv \ln(\Lambda/\Lambda_{g})$, the quartic approaches its Landau pole at $\Lambda$ very slowly as it depends on the energy scale via a double logarithm.

Another way to see that a small $|\delta|$ is likely to result in walking, is to note that in case $\delta$ is small but imaginary and taking $g^2$ constant
the one-loop $\beta$-function $\beta_{\lambda} (\lambda, g^{2})$ may have real roots which we denote by $\lambda_{\pm}$ (not to be confused with $\bar\lambda_\pm$ appearing in the definition on $\delta$, Eq.~\eqref{eq:sigma:delta}). In this simple setup $\lambda_{\pm} \propto g^{2}$ and the roots are thus evolving with $g^{2}$ when the running of $g^2$ is taken into account.  The roots $\lambda_{\pm}$ can be interpreted as \emph{pseudo-fixed points} and, given the ordering $\lambda_{+} \geqslant \lambda_{-}$, they behave qualitatively as a UV repulsor and a UV attractor respectively. In case $\lambda_{\pm}$ are real, the walking behaviour of $\lambda$ now results from the fact that the $\beta$-function of $\lambda$ can become negative when $\lambda \in (\lambda_{-},\lambda_{+})$ and thus $\lambda$ will decrease within this region.
In case of real $\bar\lambda_{\pm}$, that is when
\be
	(b_{\lambda g} - b_{g})^{2} > 4b_{\lambda}b_{\lambda gg},
	\label{eq:fixed:flow}
\ee
the quartic coupling $\lambda$ can follow the running of $g^{2}$ indefinitely. For real $\lambda_{\pm}$, that is for
\be
	b_{\lambda g} ^{2} > 4b_{\lambda}b_{\lambda gg},
	\label{eq:real:roots}
\ee
but for complex $\bar\lambda_{\pm}$, the quartic coupling still has a Landau pole, yet, as $\lambda$ is able to track the walking fixed points for a while, the Landau pole can be pushed above the Planck scale, where gravity is expected to contribute to the running.

Our analysis of this schematic model shows that, unlike asymptotic freedom, the walking does not purely follow from the field content of the model, but also depends on the initial conditions for the RGEs. It also shows that some models are more likely to exhibit walking behaviour than others. Finally, since for a real $\delta$ the RGE flow can be totally asymptotically free, thus, in the context of the condition $\Im \delta \ll \pi$, walking seems to be characteristic to closeness to total asymptotic freedom. It is interesting to remark, that the SM possesses the last property~\cite{Giudice:2014tma} as the quartic coupling of the Higgs boson shows walking behaviour.

\subsection{Running, walking and Landau poles for higher $SU(3)_c$ multiplets}

Next, we consider extending SM by a coloured scalar multiplet in representation $\mathbf{3}$, $\mathbf{6}$, $\mathbf{8}$, $\mathbf{10}$, $\mathbf{15}$ or $\mathbf{15}^\prime$.
Each representation will be considered in turn and we calculate the RGEs for all the models under study at one-loop level with the help of the PyR@TE 2 package \cite{Lyonnet:2013dna,Lyonnet:2016xiz}.\footnote{The package allows for two-loop calculation, but in the case of the $\mathbf{15}$ or $\mathbf{15'}$, even the one-loop computation takes a couple of weeks on a 3.4 GHz Core~i7 processor.} The RGEs are presented in Appendix~\ref{app:rge}. To find the correspondence between our bases for the gauge invariant contractions and the bases used in the PyR@TE, the presentation of a representation in PyR@TE can be extracted by contracting it with an appropriate number of (anti)fundamental representations to form a singlet. For example, an adjoint scalar $S_{i}^{j}$ of $SU(3)$ can be contracted with one triplet $a^{i}$ and one antitriplet $b_{j}$ to form the singlet $a^{i} S_{i}^{j} b_{j}$. The singlet can be calculated in a PyR@TE interactive session and the PyR@TE basis for the $S_{i}^{j}$ extracted as the coefficient tensor for  $a^{i}$ and $b_{j}$.

\begin{figure}[tb]
\centering
\includegraphics{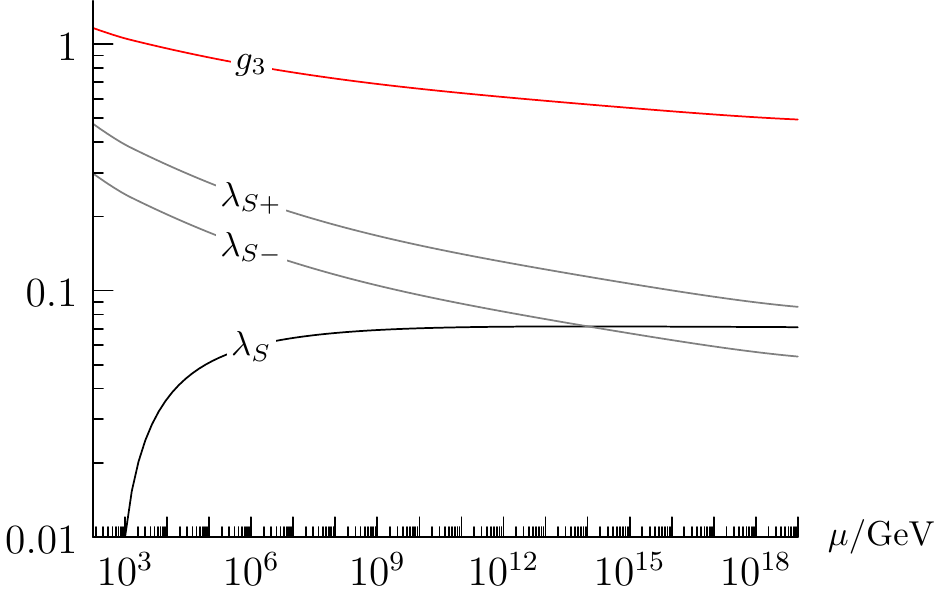}
\caption{The running of the scalar self-coupling $\lambda_{S}$ of $\mathbf{3}$ in the case where we set $\lambda_{SH} = 0$ for simplicity. The $\beta$-function $\beta_{\lambda_{S}} = 0$ at $\lambda_{S\pm}$. Between $\lambda_{S-}$ and $\lambda_{S+}$ the $\beta$-function is negative and remains small in a large energy range due to crossing zero twice. The scalar self-coupling $\lambda_{S}$ does not run, but \emph{walks}.}
\label{fig:walking}
\end{figure}

\begin{figure}[tb]
\centering
\includegraphics{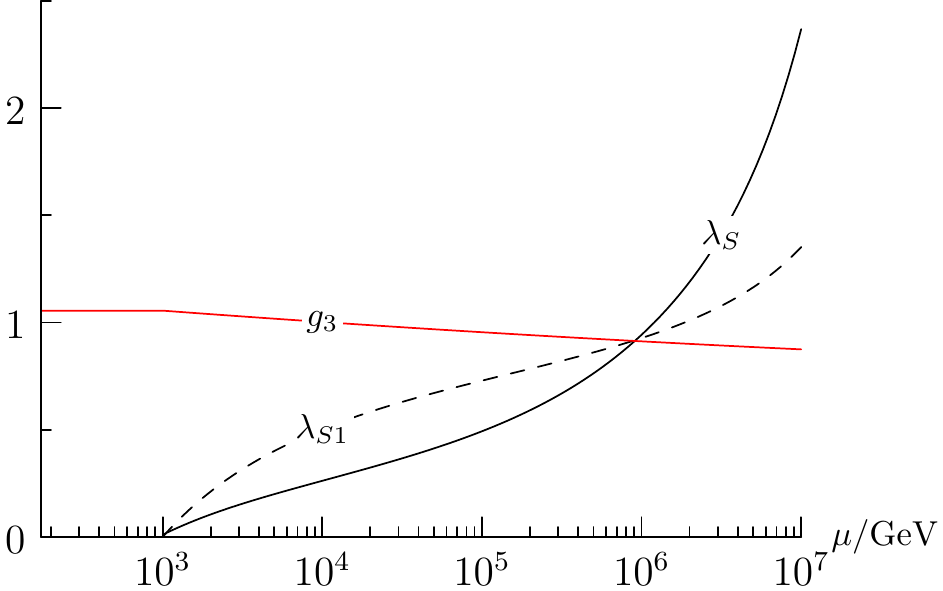}
\caption{Renormalisation group running for the representation $\mathbf{10}$ of $SU(3)$ in the case where we take $\lambda_{SH} = 0$. At 1-loop level the theory breaks down well below the Planck scale because the self-couplings $\lambda_{S}$ and $\lambda_{S1}$ develop a Landau pole due to their strong dependence on $g_3$.}
\label{fig:running:10}
\end{figure}

\begin{figure}[htbp]
\begin{center}
  \includegraphics{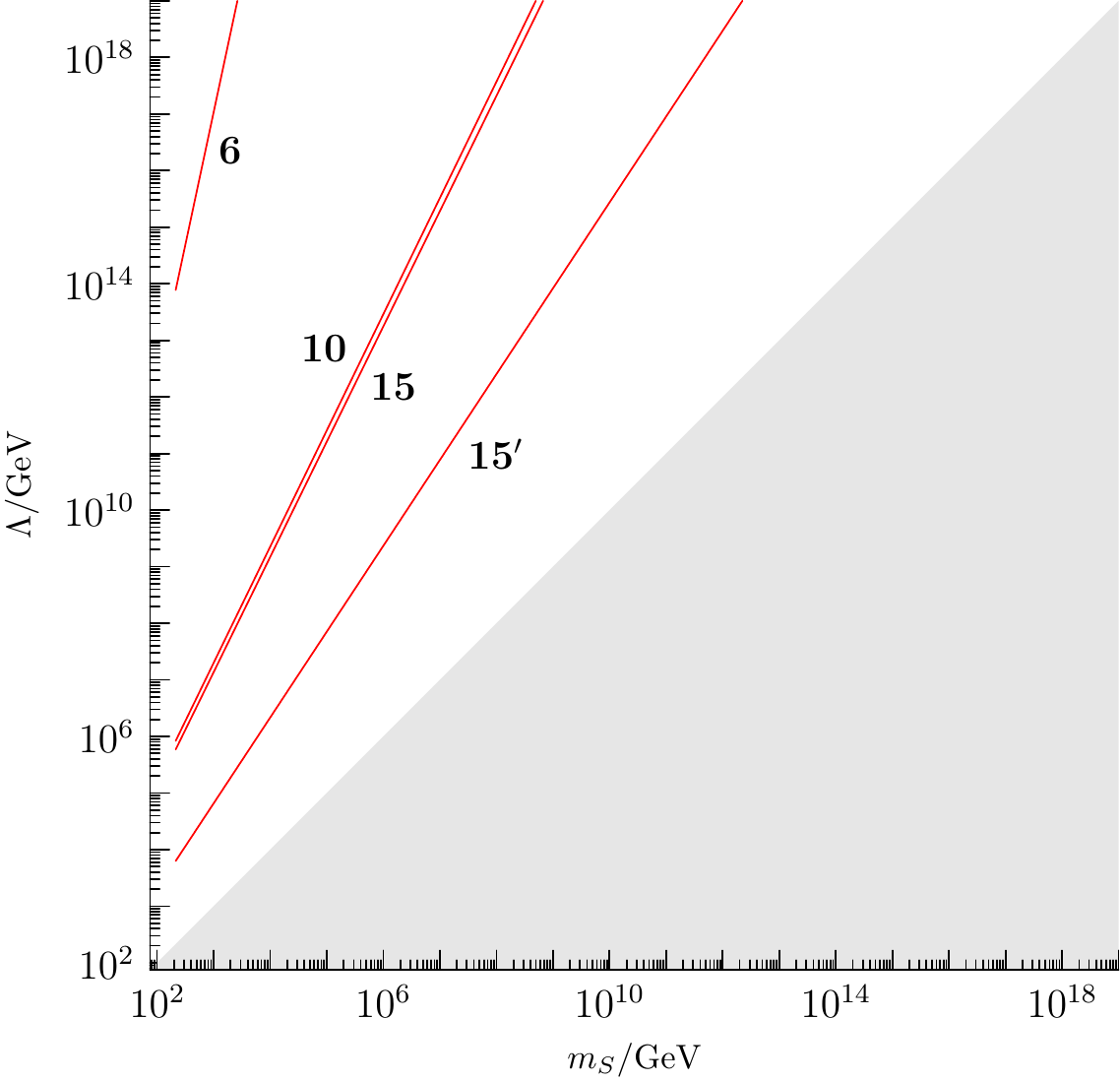}
\caption{Perturbativity bounds for SM extended by one coloured scalar multiplet $S$ as a function of $m_{S}$. Below $m_{S}$ the effective theory is given by the SM (gray). The Landau poles of the $\mathbf{3}$ and $\mathbf{8}$ (not shown) are far above the Planck scale for the depicted mass range.}
\label{fig:nonpert:QCD}
\end{center}
\end{figure}

In general, the larger the multiplet, the faster we expect the quartic scalar couplings to run. For most of the multiplets it appears to be true. The self-couplings of the representations $\mathbf{3}$ and $\mathbf{8}$, however, change very slowly for a large range of energies. Indeed, their $\beta$-functions satisfy the conditions for the existence of pseudo-fixed points given at the end of Section~\ref{sec:toy:model}.

Fig.~\ref{fig:walking} demonstrates the walking self-coupling of the fundamental representation of $SU(3)_{c}$. We set $\lambda_{S} = 0$ at $m_{S} = 1$~TeV. We set $\lambda_{SH} = 0$ for simplicity (the qualitative behaviour remains the same with a non-zero portal coupling). The roots of the $\beta_{\lambda_{S}}$ \eqref{eq:beta:lambda:S:of:3} are given by $\lambda_{S\pm} = (1/84) \, (24 \pm \sqrt{30}) \, g_{3}^{2}$ which are real.\footnote{To obtain the walking behaviour, obviously the initial value of $\lambda_{S}$ must be at least below $\lambda_{S+}$.} But $\bar\lambda_{\pm}$ defined in \eqref{ex:RGE_walking_2} are complex: their average $\sigma_{\bf 3} = 7/82$ and half-difference $\delta_{\bf 3} = 0.56 \, i$. Although $\lambda_{S}$ will eventually hit a Landau pole (due to imaginary $\delta$), this will happen at a much higher scale than the Planck scale, where quantum gravity may, arguably, solve the problem.

For multiplets other than the fundamental $\mathbf{3}$ and the adjoint $\mathbf{8}$ of $SU(3)$, the Landau poles appear much below the Planck scale. An example with the RGE running of the self-couplings of the representation $\mathbf{10}$ is shown in Fig.~\ref{fig:running:10}, where the mass $m_{S} = 1$~TeV and $\lambda_{S}(m_{S}) = \lambda_{S1}(m_{S}) = 0$. The qualitative relation between the mass and the Landau pole can be understood in the approximation of the schematic model. Ignoring other couplings, we can evaluate $\delta$ parameter \eqref{eq:sigma:delta} for the the RGE of the strong coupling $g_{s}$ and the fastest growing quartic $\lambda_{S}$. For example, for higher representations we have $\delta_{\bf 6} = 1.34 \, i$, $\delta_{\bf 10} = 1.34 \, i$, $\delta_{\bf 15'} = 14.52 \, i$. Comparing this with $\delta_{\bf 3} = 0.56 \, i$, we expect the Landau poles to appear at lower scales. However, as seen from Fig.~\ref{fig:nonpert:QCD}, this oversimplified approach clearly fails to capture the difference between representations $\bf 6$ and $\bf 10$ or the similarity between $\bf 10$ and $\bf 15$. A more careful analysis could be based on the study of asymptotics of the RGEs outlined e.g. in Ref.~\cite{Giudice:2014tma}. This lies, however, beyond the scope of this paper.

In Fig.~\ref{fig:nonpert:QCD} we show the scales of the Landau poles as a function of the mass of the multiplet $S$. In the gray triangle the energy scale is below $m_{S}$, so the effective theory is the SM. The Landau poles for the representations $\mathbf{3}$ and $\mathbf{8}$ are much higher, at about $10^{42}$~GeV, and are not shown. The quartic self-couplings of all multiplets can be made perturbative up to the Planck scale, provided that the mass of $S$ is chosen to be high enough.\footnote{In case of a dark $SU(N)$, of course, the Landau poles can be pushed up by making the gauge coupling small.}

The vacuum stability conditions allow for some quartic scalar self-couplings to be negative. Nevertheless, we find that $\lambda_{S} < 0$ either does not noticeably change the position of the Landau poles or bring them lower. In the presence of Yukawa couplings, however, a negative $\lambda_{S}$ can be crucial to establish an asymptotically safe solution \cite{Giudice:2014tma}. Then it is important to know the exact vacuum stability conditions to check whether the potential is bounded from below for the given solution.

\subsection{New scales from strong dynamics of higher scalar multiplets}

A generic feature in strongly coupled theories is the appearance of different condensation scales for different multiplets~\cite{Marciano:1980zf,Lust:1985aw,Kubo:2014ova}. In the SM with a scalar $S$ in a higher representation $\mathbf{R}$ of colour, a new QCD scale larger than the usual $\Lambda_{\text{QCD}}$ could appear. The confinement of the scalar $S$ takes place at the scale $\Lambda_s$ determined by
\be
C_2(\mathbf{R}) \alpha_s(\Lambda_s)\gsim \kappa={\cal O}(1),
\label{eq:condens}
\ee
where $C_2(\mathbf{R})$ is the Casimir of the scalar representation $\mathbf{R}$.
For sufficiently large $C_2(\mathbf{R})$ the condensate $ \langle\bar SS\rangle$ forms for perturbative coupling $\alpha_s$ e.g. at $\mu = 10$~GeV for the $\mathbf{10}$,  at $\mu = 6$~GeV for the $\mathbf{15}$, and  at $\mu = 195$~GeV for the $\mathbf{15'}$ for $\kappa = 1$ according to Eq.~\eqref{eq:condens} and the values in Table~\ref{table:reps}.

It is expected that also the quartic self-couplings would give a contribution to the left hand side of Eq.~\eqref{eq:condens}. The problem with quartics, however, is that their Landau poles lie in the UV, not in the IR as for the strong gauge coupling. Therefore, Landau poles of quartic self-couplings of $S$ at sufficiently low scale will completely invalidate the analysis of the condensation mechanism.

New interactions, such as Yukawa couplings, that give negative contributions to the $\beta$-functions of scalar quartics can be introduced to remove Landau poles. But bringing in new Dirac fermions to produce the required Yukawas would only work for lower QCD multiplets, because additional fermions may endanger the asymptotic freedom of the gauge coupling.

For QCD with $\kappa = 1$, the scales generated by any higher multiplets are too low to be compatible with current experimental limits. For these reasons, one cannot use large scalar QCD multiplets for a dynamical generation of a new mass scale as was attempted in~\cite{Kubo:2014ova}. The mechanism could be used, however, for dark $SU(N)$ gauge groups where the number of colours can be adjusted and on which the experimental bounds are far more lenient than on QCD.

\section{Conclusions}
\label{sec:out}

We studied the vacuum stability and RGE running of the SM extended by one higher scalar colour multiplet that preserves the asymptotic freedom of the strong gauge coupling. We derived the bounded-from-below conditions for the scalar potential and studied the Landau poles arising from the running scalar quartic couplings. For both, we presented a general analysis before embarking on the study of specific models. The conditions that result from our analysis can be imposed on extensions of the SM or on models where a dark $SU(3)$ is used, e.g., for flavour or for dark matter.

In order to derive the vacuum stability conditions, we studied the orbit spaces of the quartic self-coupling terms of the multiplets. If the scalar potential depends on the orbit space parameters linearly, the vacuum stability conditions are determined by the convex hull of the orbit space. For most of the multiplets we found simple analytical expression for the necessary and sufficient vacuum stability conditions. For the representation $\mathbf{15}$, the orbit space is 4-dimensional and more complicated, and we determined the convex hull of its orbit space numerically.

The running of the self-couplings of the $\mathbf{6}$, $\mathbf{10}$ and $\mathbf{15}$ can be made perturbative up to the Planck scale if the mass of the scalar is set high enough. It is not possible for the $\mathbf{15'}$ whose scalar self-couplings immediately hit the Landau pole.
On the other hand, the scalar self-couplings $\lambda_{S}$ for the $\mathbf{3}$ and $\mathbf{8}$ multiplets \emph{walk} rather than run: they maintain perturbativity up to scales of about $10^{42}$~GeV.  This effect -- similar to the running of the Higgs coupling $\lambda_{H}$ in the SM -- occurs because the running $\lambda_{S}$ stays near the zeroes of its $\beta$-function. We presented a generic description of walking quartic couplings in terms of pseudo-fixed points. The RGE analysis we present gives a conservative estimate also for models where the multiplets have additional gauge charges, because this, in most cases, will only bring the Landau poles lower.

We also studied the possibility of generating new high scales from the strong dynamics of higher scalar coloured multiplets. Taking into account all the constraints,
the scales produced are too low to provide the origin of the electroweak scale. However, in models beyond the SM in which the $SU(3)$ gauge group is not associated with colour,
this mechanism may be used to generate interesting phenomenology. Our results are also applicable to model building utilising any $SU(N)$ and scalars in fundamental
or higher representations.


\section*{Acknowledgments}
We would like to thank Christian Gross, Manfred Lindner and Renato Fonseca for useful discussions. This work was supported by the Estonian Research Council grant PUT799, the grant IUT23-6 of the Estonian Ministry of Education and Research, Academy of Finland grant 267842, and by the EU through the ERDF CoE program grant TK133. The work of F.L. was also partially supported by the U.S. Department of Energy under Grant No. DESC0010129.

\appendix

\section{Renormalisation Group Equations}
\label{app:rge}

We have calculated the RGEs with the help of the PyR@TE package \cite{Lyonnet:2013dna,Lyonnet:2016xiz}. We present the RGEs at 1-loop level for the SM extended with one scalar with the quantum numbers
 $S(\mathbf{R},1,0)$. The RGEs for the self-couplings of $\mathbf{3}$, $\mathbf{6}$ and $\mathbf{8}$ have been derived previously in \cite{Gross:1973ju} for general $SU(N)$ gauge theories without the Higgs portal terms. The RGEs output by PyR@TE agree with \cite{Gross:1973ju} in full for the $\mathbf{3}$, $\mathbf{6}$ and for the $\lambda_{S}^{2}$ and $\lambda_{S} g_{3}^{3}$ terms for the $\mathbf{8}$.

\subsection{RGEs for SM and $S$ in $\mathbf{3}$}

\begin{align}
16 \pi^2 \beta_{g_Y} &= \frac{41 g_Y^3}{6}, \\
16 \pi^2 \beta_{g} &= -\frac{19 g^3}{6}, \\
16 \pi^2 \beta_{g_3} &= -\frac{41 g_3^3}{6}, \\
  16 \pi^{2} \beta_{y_{t}} &= y_{t} \left( \frac{9}{2} y_{t}^{2} - \frac{17}{12} g^{\prime 2}
  - \frac{9}{4} g^{2} - 8 g_{3}^{2} \right),  \\
16 \pi^2 \beta_{\lambda_H} &= \frac{9 g^4}{8}+\lambda_H \left(-9 g^2-3 g_Y^2+12 y_t^2\right)+\frac{3}{4} g^2 g_Y^2+\frac{3 g_Y^4}{8}+24 \lambda_H^2 \notag \\
&+ 3 \lambda_{SH}^2-6 y_t^4, \\
16 \pi^2 \beta_{\lambda_S} &= -16 g_3^2 \lambda_S+\frac{13 g_3^4}{6}+28 \lambda_S^2+2 \lambda_{SH}^2,
  \label{eq:beta:lambda:S:of:3} \\
16 \pi^2 \beta_{\lambda_{SH}} &= \lambda_{SH} \left(-\frac{9 g^2}{2}-\frac{3 g_Y^2}{2}-8 g_3^2+12 \lambda_H+16 \lambda_S+6 y_t^2\right)+4 \lambda_{SH}^2.
\end{align}

\subsection{RGEs for SM and $S$ in $\mathbf{6}$}

\begin{align}
16 \pi^2 \beta_{g_Y} &= \frac{41 g_Y^3}{6}, \\
16 \pi^2 \beta_{g} &= -\frac{19 g^3}{6}, \\
16 \pi^2 \beta_{g_3} &= -\frac{37 g_3^3}{6}, \\
  16 \pi^{2} \beta_{y_{t}} &= y_{t} \left( \frac{9}{2} y_{t}^{2} - \frac{17}{12} g^{\prime 2}
  - \frac{9}{4} g^{2} - 8 g_{3}^{2} \right),  \\
16 \pi^2 \beta_{\lambda_H} &= \frac{9 g^4}{8}+\lambda_H \left(-9 g^2-3 g_Y^2+12 y_t^2\right)+\frac{3}{4} g^2 g_Y^2+\frac{3 g_Y^4}{8}+24 \lambda_H^2 \notag \\
&+ 6 \lambda_{SH}^2-6 y_t^4, \\
16 \pi^2 \beta_{\lambda_S} &= \left(32 \lambda_{S1}-40 g_3^2\right) \lambda_S+\frac{35 g_3^4}{3}+6 \lambda_{S1}^{2}+40 \lambda_S^2+2 \lambda_{SH}^2, \\
16 \pi^2 \beta_{\lambda_{S1}} &= \lambda_{S1} \left(24 \lambda_S-40 g_3^2\right)+5 g_3^4+22 \lambda_{S1}^{2}, \\
16 \pi^2 \beta_{\lambda_{SH}} &= \lambda_{SH} \left(-\frac{9 g^2}{2}-\frac{3 g_Y^2}{2}-20 g_3^2+12 \lambda_H+16 \lambda_{S1}+28 \lambda_S+6 y_t^2\right)+4 \lambda_{SH}^2.
\end{align}

\vspace{2cm}

\subsection{RGEs for SM and $S$ in $\mathbf{8}$}

\begin{align}
16 \pi^2 \beta_{g_Y} &= \frac{41 g_Y^3}{6}, \\
16 \pi^2 \beta_{g} &= -\frac{19 g^3}{6}, \\
16 \pi^2 \beta_{g_3} &= -6 g_3^3, \\
  16 \pi^{2} \beta_{y_{t}} &= y_{t} \left( \frac{9}{2} y_{t}^{2} - \frac{17}{12} g^{\prime 2}
  - \frac{9}{4} g^{2} - 8 g_{3}^{2} \right), \\
  16 \pi^{2} \beta_{\lambda_{H}} &= 24 \lambda_H^2 + \lambda_H (12 y_t^2 - 3 g^{\prime 2} - 9 g^2)
  + \frac{3}{8} (g^{\prime 4} + 2 g^{\prime 2} g^2 +3 g^4) \notag \\
  &- 6 y_t^4 + 4 \lambda_{SH}^2 \\
    16 \pi^2 \beta_{\lambda_S} &= 32 \lambda_S^2 -36 g_3^2 \lambda_S+\frac{9}{16} g_3^4 + 2 \lambda_{SH}^2, \\
  16 \pi^{2} \beta_{\lambda_{SH}} &= \lambda_{SH} \left( 20 \lambda_S + 12 \lambda_H + 6 y_t^2 - \frac{3}{2} g^{\prime 2} - \frac{9}{2} g^2 - 18 g_3^2 \right) + 4 \lambda_{SH}^2.
\end{align}

\subsection{RGEs for SM and $S$ in $\mathbf{10}$}

 \begin{align}
  16 \pi^{2} \beta_{g'} &= \frac{41}{6} g^{\prime 3}, 
  \\
  16 \pi^{2} \beta_{g} &= -\frac{19}{6} g^{3}, 
  \\
  16 \pi^{2} \beta_{g_{3}} &= -\frac{9}{2} g_{3}^{3}, 
  \\
  16 \pi^{2} \beta_{y_{t}} &= y_{t} \left( \frac{9}{2} y_{t}^{2} - \frac{17}{12} g^{\prime 2}
  - \frac{9}{4} g^{2} - 8 g_{3}^{2} \right),  
  \\
  16 \pi^{2} \beta_{\lambda_{H}} &= 24 \lambda_H^2 + \lambda_H (12 y_t^2 - 3 g^{\prime 2} - 9 g^2)
  + \frac{3}{8} (g^{\prime 4} + 2 g^{\prime 2} g^2 +3 g^4) \notag \\
  &- 6 y_t^4 + 10 \lambda_{SH}^2,
  \\
  16 \pi^{2} \beta_{\lambda_{S}} &= 56 \lambda_S^2 + \frac{32}{9} \lambda_{S1}^{2} + \lambda_S \left( 40 \lambda_{S1} - 72 g_3^2 \right) + \frac{81}{4} g_3^4 + 2 \lambda_{SH}^2,
    \\
  16 \pi^{2} \beta_{\lambda_{S1}} &= \frac{232}{9} \lambda_{S1}^{2}
  + \lambda_{S1} \left( 24 \lambda_{S} - 72 g_{3}^{2} \right) + \frac{189}{4} g_{3}^{4},
  \\
  16 \pi^{2} \beta_{\lambda_{SH}} &= \lambda_{SH} \left( 44 \lambda_S + 20 \lambda_{S1} + 12 \lambda_H + 6 y_t^2 - \frac{3}{2} g^{\prime 2} - \frac{9}{2} g^2 - 36 g_3^2 \right) + 4 \lambda_{SH}^2.
\end{align}

\vspace{2cm}

\subsection{RGEs for SM and $S$ in $\mathbf{15}$}

\begin{align}
  16 \pi^2 \beta_{g_Y} &= \frac{41}{6} g_Y^3, \\
  16 \pi^2 \beta_{g} &= -\frac{19}{6} g^3, \\
  16 \pi^2 \beta_{g_3} &= -\frac{11}{3} g_3^3, \\
  16 \pi^{2} \beta_{y_{t}} &= y_{t} \left( \frac{9}{2} y_{t}^{2} - \frac{17}{12} g^{\prime 2}
  - \frac{9}{4} g^{2} - 8 g_{3}^{2} \right), \\
  16 \pi^2 \beta_{\lambda_H} &= \frac{9 g^4}{8}+\lambda_H \left(12 y_t^2 -9 g^2 - 3 g_Y^2 \right)+\frac{3}{4} g^2
  g_Y^2+\frac{3 g_Y^4}{8}+24 \lambda_H^2 \notag \\
  & -6 y_t^4 +15 \lambda_{SH}^2, \\
  16 \pi^2 \beta_{\lambda_S} &= \lambda_S \left(-64 g_3^2+55 \lambda_{S1}+38 \lambda_{S2}
  +60 \lambda_{S3}+18 \lambda_{S4}\right)-\frac{61}{12} g_3^4
  + \frac{115}{8} \lambda_{S1} \lambda_{S2} \notag \\
  &+13 \lambda_{S1} \lambda_{S3}+\frac{9}{4} \lambda_{S1} \lambda_{S4}+\frac{87}{16} \lambda_{S1}^2+\frac{85}{8} \lambda_{S2} \lambda_{S3}+\frac{11}{8} \lambda_{S2} \lambda_{S4}+\frac{317}{64} \lambda_{S2}^{2}
  \notag \\
  &+\frac{1}{2} \lambda_{S3} \lambda_{S4} +\frac{57}{4} \lambda_{S3}^{2} +\lambda_{S4}^{2}+76 \lambda_S^2+2 \lambda_{SH}^2, \\
  16 \pi^2 \beta_{\lambda_{S1}} &= \lambda_{S1} \left(-64 g_3^2-\frac{27 \lambda_{S2}}{4}+35 \lambda_{S3}+24 \lambda_S\right)+29 g_3^4+\frac{111}{4} \lambda_{S1}^2
  -\frac{5}{2} \lambda_{S2} \lambda_{S3}
   \notag \\
  &+8 \lambda_{S2} \lambda_{S4}+\frac{49}{16} \lambda_{S2}^{2}+\lambda_{S3}^{2} + 6 \lambda_{S4}^{2}, \\
16 \pi^2 \beta_{\lambda_{S2}} &= \lambda_{S2} \left(-64 g_3^2+\frac{139 \lambda_{S1}}{8}+30 \lambda_{S3}+\frac{19 \lambda_{S4}}{4}+24 \lambda_S\right)+43 g_3^4-\frac{1}{2} \lambda_{S1} \lambda_{S3} \notag \\
  &+\frac{27}{2} \lambda_{S1} \lambda_{S4}-\frac{1}{2} \lambda_{S1}^2+\frac{51}{16} \lambda_{S2}^{2}+25 \lambda_{S3} \lambda_{S4}-11 \lambda_{S3}^{2}+2 \lambda_{S4}^{2}, \\
16 \pi^2 \beta_{\lambda_{S3}} &= \lambda_{S3} \left(-64 g_3^2+18 \lambda_{S1}-\frac{35 \lambda_{S2}}{8}+\frac{\lambda_{S4}}{2}+24 \lambda_S\right)+\frac{77 g_3^4}{4}-\frac{5}{2} \lambda_{S1} \lambda_{S2} \notag \\
& +\frac{5}{2} \lambda_{S1}^2 + \frac{45}{8} \lambda_{S2} \lambda_{S4}+\frac{233}{64} \lambda_{S2}^{2}+\frac{97 \lambda_{S3}^{2}}{4}+\frac{\lambda_{S4}^{2}}{4}, \\
  16 \pi^2 \beta_{\lambda_{S4}} &= \frac{49}{4} \lambda_{S4}^{2} + \lambda_{S4}
  \left(-64 g_3^2+\frac{39 \lambda_{S1}}{4}+\frac{31 \lambda_{S2}}{2}-\lambda_{S3}+24 \lambda_S\right)+\frac{9}{4} \lambda_{S1} \lambda_{S2}
  \notag \\
  & -\frac{7}{2} \lambda_{S1} \lambda_{S3} - \frac{7}{16} \lambda_{S1}^{2} - 3 \lambda_{S2} \lambda_{S3}+\frac{11}{4} \lambda_{S2}^{2}+\lambda_{S3}^{2}, \\
16 \pi^2 \beta_{\lambda_{SH}} &= \lambda_{SH} \left(12 \lambda_H + 64 \lambda_S + \frac{55}{2} \lambda_{S1} + 19 \lambda_{S2}+30 \lambda_{S3}+9 \lambda_{S4}+6 y_t^2
  \right.
\notag \\
 & \left. -\frac{9 g^2}{2}-\frac{3 g_Y^2}{2}-32 g_3^2 \right)+4 \lambda_{SH}^2.
\end{align}

\vspace{2cm}

\subsection{RGEs for SM and $S$ in $\mathbf{15'}$}

\begin{align}
  16 \pi^{2} \beta_{g'} &= \frac{41}{6} g^{\prime 3}, 
  \\
  16 \pi^{2} \beta_{g} &= -\frac{19}{6} g^{3}, 
  \\
  16 \pi^{2} \beta_{g_{3}} &= -\frac{7}{6} g_{3}^{3}, 
  \\
  16 \pi^{2} \beta_{y_{t}} &= y_{t} \left( \frac{9}{2} y_{t}^{2} - \frac{17}{12} g^{\prime 2}
  - \frac{9}{4} g^{2} - 8 g_{3}^{2} \right), 
  \\
  16 \pi^{2} \beta_{\lambda_{H}} &= 24 \lambda_{H}^2 + \lambda_H (12 y_t^2 - 3 g^{\prime 2} - 9 g^2) + \frac{3}{8} (g^{\prime 4} + 2 g^2 g^{\prime 2}+ 3 g^4) \notag \\
  & - 6 y_{t}^4 + 15 \lambda_{SH}^2,
  \\
  16 \pi^{2} \beta_{\lambda_{S}} &= 76 \lambda_{S}^{2} + \lambda_{S} (52 \lambda_{S1} + 40 \lambda_{S2} -112 g_{3}^2) + 2 \lambda_{SH}^2 + \frac{9}{2} \lambda_{S1}^{2} + 4 \lambda_{S1} \lambda_{S2} \notag \\
  &+ \frac{2}{3} \lambda_{S2}^{2} + \frac{164}{3} g_{3}^4,
    \\
  16 \pi^{2} \beta_{\lambda_{S1}} &= 23 \lambda_{S1}^{2} + \lambda_{S1} (-112 g_{3}^2 + 24 \lambda_{S} +
    36 \lambda_{S2}) +
 \frac{128}{9} \lambda_{S2}^{2} -76 g_{3}^4,
  \\
  16 \pi^{2} \beta_{\lambda_{S2}} &= \frac{130}{9} \lambda_{S2}^{2} + \lambda_{S2} \left(24 \lambda_{S} +
    20 \lambda_{S1} -112 g_{3}^2 \right) + \frac{9}{2} \lambda_{S1}^{2} + 216 g_{3}^4,
  \\
  16 \pi^{2} \beta_{\lambda_{SH}} &= \lambda_{SH} \left(6 y_{t}^2 + 12 \lambda_{H} + 64 \lambda_{S} + 26 \lambda_{S1} +  20 \lambda_{S2} - \frac{3}{2} g^{\prime 2} -\frac{9}{2}g^2 - 56 g_{3}^2 \right)
  \notag \\
  &+ 4 \lambda_{SH}^{2}.
\end{align}

\section{Bases for higher representations}
\label{sec:appB}

\subsection{Basis of the $\mathbf{15'}$ of $SU(3)$}
\label{sec:param:15p}

For the representation $\mathbf{15'}$ of $SU(3)$, we use the basis $a_{i}$ with $i = 1, \ldots, 15$. The independent elements of the tensor $S^{ijkl}$ are given by
\begin{align}
	S^{1111} &= a_{1},
	&
	S^{1112} &= \frac{1}{2} a_{2},
	&
	S^{1113} &= \frac{1}{2} a_{3},
	&
	S^{1122} &= \frac{1}{\sqrt 6} a_{4},
	\\
	S^{1123} &= \frac{1}{2 \sqrt 3} a_{5},
	&
	S^{1133} &= \frac{1}{\sqrt 6} a_{6},
	&
	S^{1222} &= \frac{1}{2} a_{7},
	&
	S^{1223} &= \frac{1}{2\sqrt 3} a_{8},
	\\
	S^{1233} &= \frac{1}{2\sqrt 3}  a_{9},
	&
	S^{1333} &= \frac{1}{2}  a_{10},
	&
	S^{2222} &= a_{11},
	&
	S^{2223} &= \frac{1}{2}  a_{12},
	\\
	S^{2233} &= \frac{1}{\sqrt 6}  a_{13},
	&
	S^{2333} &= \frac{1}{2} a_{14},
	&
	S^{3333} &= a_{15}.
\end{align}
This defines an orthogonal basis with respect to the norm $S^{ijkl}S_{ijkl}$.

\subsection{Basis of the $\mathbf{15}$ of $SU(3)$}
\label{sec:param:15}

For the representation $\mathbf{15}$ of $SU(3)$, we use the basis $a_{i}$ with $i = 1, \ldots, 15$. The independent elements of the tensor $S^{ij}_{k}$ are given by
\begin{align}
  S^{11}_{1} &= \frac{1}{\sqrt{6}} (-\sqrt{2} a_{4} + a_{5}),
  &
  S^{11}_{2} &= a_{2},
  &
  S^{11}_{3} &= -a_{1},
  \\
  S^{12}_{1} &= \frac{1}{2 \sqrt{6}} (-2 \sqrt{2} a_{8} + a_{9}),
  &
  S^{12}_{2} &= \frac{1}{2 \sqrt{6}} (2 \sqrt{2} a_{4} + a_{5}),
  &
  S^{12}_{3} &= -\frac{1}{\sqrt{2}} a_{3},
  \\
  S^{13}_{1} &= \frac{1}{4} (-2 a_{10} + \sqrt{2} a_{11}),
  &
  S^{13}_{2} &= \frac{1}{\sqrt{2}} a_{6},
  &
  S^{13}_{3} &= -\frac{1}{2} \sqrt{\frac{3}{2}} a_{5},
  \\
  S^{22}_{1} &= -a_{13},
  &
  S^{22}_{2} &= \frac{1}{\sqrt{6}} (\sqrt{2} a_{8} + a_{9}),
  &
  S^{22}_{3} &= -a_{7},
  \\
  S^{23}_{1} &= -\frac{1}{\sqrt{2}} a_{14},
  &
  S^{23}_{2} &= \frac{1}{4} (2 a_{10} +\sqrt{2} a_{11}),
  &
  S^{23}_{3} &= -\frac{1}{2} \sqrt{\frac{3}{2}} a_{9},
  \\
  S^{33}_{1} &= -a_{15},
  &
  S^{33}_{2} &= a_{12},
  &
  S^{33}_{3} &= -\frac{1}{\sqrt{2}} a_{11}.
\end{align}
This defines an orthogonal basis with respect to the norm $S^{ij}_{k}S_{ij}^{k}$.

\bibliographystyle{JHEP}
\bibliography{QCDscalars}

\end{document}